\documentclass[aps,prl,reprint,superscriptaddress,showpacs]{revtex4-1}

\usepackage{graphicx}
  \setkeys{Gin}{width=\linewidth}
\usepackage{amsmath,amsfonts,amssymb}

\begin{document}


\title{Optical selection rules of graphene nanoribbons}


\author{H. C. Chung}
\affiliation{Department of Physics, National Cheng Kung University, Tainan 70101, Taiwan}
\author{M. H. Lee}
\affiliation{Department of Physics, National Cheng Kung University, Tainan 70101, Taiwan}
\author{C. P. Chang}
\email[]{t00252@mail.tut.edu.tw}
\affiliation{Center for General Education, Tainan University of Technology, 710 Tainan, Taiwan}
\author{M. F. Lin}
\email[]{mflin@mail.ncku.edu.tw}
\affiliation{Department of Physics, National Cheng Kung University, Tainan 70101, Taiwan}


\date{\today}

\begin{abstract}
Optical selection rules for one-dimensional graphene nanoribbons are analytically studied and clarified based on the tight-binding model.
A theoretical explanation, through analyzing the velocity matrix elements and the features of wavefunctions, can account for the selection rules, which depend on the edge structure of nanoribbon, namely armchair or zigzag edges.
The selection rule of armchair nanoribbons is $\Delta J=J^{c}-J^{v}=0$, and the optical transitions occur from the conduction to valence subbands of the same index.
Such a selection rule originates in the relationships between two sublattices and between conduction and valence subbands.
On the other hand, zigzag nanoribbons exhibit the selection rule $|\Delta J|=odd$, which results from the alternatively changing symmetry property as the subband index increases.
An efficiently theoretical prediction on transition energies is obtained with the application of selection rules.
Furthermore, the energies of band edge states become experimentally attainable via optical measurements.
\end{abstract}


\maketitle

\section{Introduction}

The discovery of graphene~\cite{Science306(2004)666K.S.Novoselov, PNAS102(2005)10451K.S.Novoselov} has motivated many studies because of its interesting physical properties and possible applications.
Graphene, a two-dimensional carbon-based material~\cite{Science324(2009)875R.F.Service, Europhys.News40.3(2009)17N.M.R.Peres, MaterialsToday10(2007)20M.I.Katsnelson}, is a gapless semiconductor with linear energy dispersion near the Dirac point~\cite{Science324(2009)1530A.K.Geim, NatureMaterials6(2007)183A.K.Geim}.
This cone-like energy spectrum results in particular electronic and transport properties, such as massless Dirac fermions which travel at a extremely high constant speed~\cite{Phys.Rev.Lett.100(2008)016602S.V.Morozov, NatureMaterials6(2007)652F.Schedin}, finite conductivity at zero charge-carrier concentration~\cite{Physics-Uspekhi51(2008)744S.V.Morozov, Phys.Rev.B77(2008)081402S.Cho}, and unusual quantum Hall effect~\cite{Nature438(2005)197K.S.Novoselov, Nature438(2005)201Y.B.Zhang}.
However, the gapless feature of graphene limits its applications in electronic devices.
In order to induce a tunable band gap for a wide variety of applications, one-dimensional (1D) graphene nanoribbons (GNRs) are proposed, which can be fabricated using a nanowire etch mask~\cite{NanoLett.9(2009)2083J.W.Bai, Phys.StatusSolidiB246(2009)2514A.Fasoli}, chemical vapor deposition~\cite{Phys.Rev.B81(2010)245428V.L.J.Joly, Chem.Phys.Lett.469(2009)177J.Campos-Delgado}, lithographic patterning of graphene, \cite{NatureNanotechnology3(2008)397L.Tapaszto, Phys.Rev.Lett.98(2007)206805M.Y.Han, Science312(2006)1191C.Berger}, and unzipping of nanotube~\cite{Nature458(2009)872D.V.Kosynkin, Carbon48(2010)2596F.Cataldo, NanoLett.10(2010)1764M.C.Pavia, NanoLett.9(2009)1527A.G.Cano-Marquez}.
The electronic and transport properties are highly related to edge structure and ribbon width.
There are two basic types of GNRs, armchair (AGNR) and zigzag (ZGNR) ones, according to their edge structures along the longitudinal direction.
ZGNR is gapless because of the peculiar partial flat bands at the Fermi level ($E_{F}=0$), while AGNR can be either semiconducting or gapless depending on the ribbon width.
Owing to these special electronic properties, GNR has triggered many studies, e.g., geometrical and energy band structures~\cite{Appl.Surf.Sci.256(2010)5776R.H.Miwa, J.Mater.Chem.20(2010)8207S.Dutta}, as well as magnetic~\cite{PhysicaE42(2010)711H.C.Chung, J.Phys.Conf.Ser.200(2010)062015T.Nomura, Phys.Rev.B81(2010)245428V.L.J.Joly, NatureNanotechnology5(2010)655J.W.Bai, J.Phys.Soc.Jpn.80(2011)044602H.C.Chung}, transport~\cite{NewJ.Phys.11(2009)095004A.Cresti, Eur.Phys.J.B72(2009)203Y.O.Klymenko, Phys.Rev.B82(2010)035446E.Perfetto}, and optical properties~\cite{Phys.Rev.Lett.101(2008)246803J.Jiang, J.Phys.Soc.Jpn.69(2000)3529M.F.Lin}.

Optical measurements are very useful in understanding the electronic structure of a material.
Unlike the featureless absorption spectrum of graphene at low energy~\cite{NewJ.Phys.12(2010)083060C.W.Chiu}, the optical spectra of GNRs exhibit many sharp peaks relating to van Hove singularities~\cite{Phys.Rev.89(1953)1189L.VanHove, PhysicsReports431(2006)261E.B.Barros} in the density of electronic states.
These peaks satisfy certain selection rules, which depend on the edge structure.
Absorption selection rules of GNRs, which crucially decide the possible transitions, have been studied by Lin \emph{et. al.} firstly~\cite{J.Phys.Soc.Jpn.69(2000)3529M.F.Lin}.
Although these rules are described, the origin is still not verified.
In this work, the selection rules are analytically studied and a theoretical explanation is given.

\section{Tight-binding model and electronic properties}

\begin{figure}
\begin{center}
  \includegraphics[]{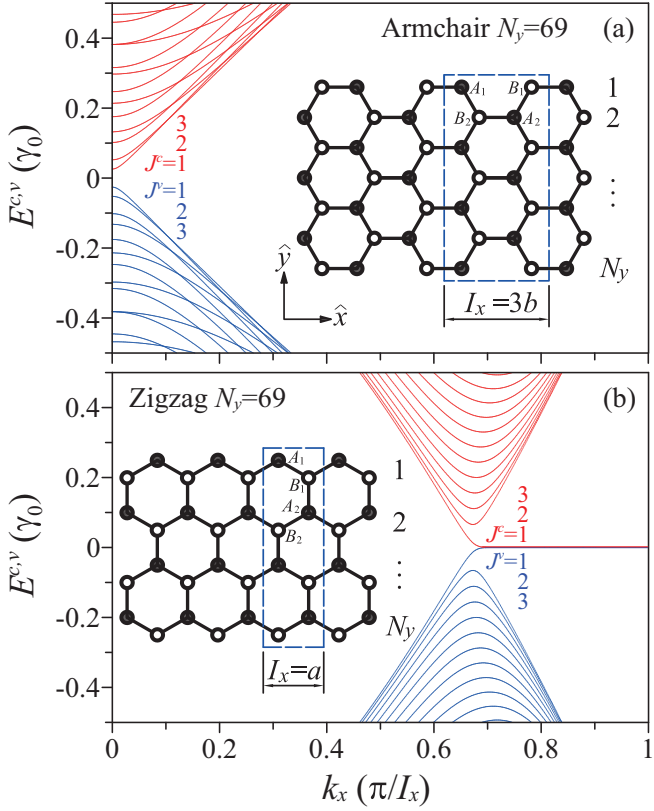}\\
  \caption{
Energy dispersions and geometric structures for $N_{y}=69$ AGNR and ZGNR.
The conduction (valence) subbands of the corresponding indices $J^{c}$ ($J^{v}$) are depicted in red (blue) color.
The dashed-line rectangles represent primitive unit cells.
The numbers of $m$th dimer line are on the right side of the hexagonal lattice.
The notations in the rectangles denote the $A$ and $B$ atoms on the $m$th dimer line.}
  \label{fig:GeometryAndBS}
\end{center}
\end{figure}

The geometric structures of AGNR and ZGNR are shown in Fig.~\ref{fig:GeometryAndBS}.
Atoms $A$ (dots) and $B$ (circles) are the nearest neighbors to each other with the C--C bond length $b=1.42$ \AA .
The ribbon width is characterized by the number ($N_{y}$) of dimer lines (armchair or zigzag lines) along the longitudinal direction ($\hat{x}$).
The periodic length is $I_{x}=3b$ ($I_{x}=\sqrt{3}b=a$) for AGNRs (ZGNRs).
The first Brillouin zone is confined in the region $-\pi /I_{x} \leq k_{x} \leq \pi /I_{x}$ with $2N_{y}$ carbon atoms in a primitive unit cell.
The Hamiltonian equation of the system is $H|\Psi\rangle = E|\Psi\rangle$.
With the framework of the tight-binding model considering only the nearest-neighbor interactions, the Hamiltonian operator is
$H=\sum_{l,l^{\prime}}\gamma_{0}c_{l}^{\dagger}c_{l^{\prime }}$, where $\gamma _{0}$ is the hopping integral, $c_{l}^{\dagger }$ ($c_{l^{\prime}}$) is the creation (annihilation) operator on the site $l$ ($l^{\prime }$), and the on-site energies are set to zero.
The Bloch wavefunction, i.e., the superposition of the $2N_{y}$ tight-binding functions $|\phi_{l}\rangle$'s with the corresponding site amplitudes $C_{l}$'s, is expressed as
\begin{equation}
|\Psi\rangle = \sum_{l=1}^{2N_{y}}C_{l} |\phi_{l}\rangle
\text{~~or~~}
|\Psi\rangle = \sum_{m=1}^{N_{y}}A_{m} |a_{m}\rangle + \sum_{m=1}^{N_{y}}B_{m} |b_{m}\rangle ,
\label{eq:WF}
\end{equation}
where $A_{m}$ ($B_{m}$) is the site amplitude and $|a_{m}\rangle$ ($|b_{m}\rangle$) is the tight-binding function associated with the periodic $A$ ($B$) atoms.

The Hamiltonian $H$ in the subspace spanned by the tight-binding functions is a $2N_{y}\times \,2N_{y}$ Hermitian matrix.
For AGNRs, the upper triangular elements of $H$ are
\begin{equation}
H_{l,l^{\prime }}=\left\{
\begin{array}{ l l l }
\gamma _{0}\exp [ik_{x}(-b)],    & \text{if } l^{\prime }=l+1, & l \text{ is odd}, \\
\gamma _{0}\exp [ik_{x}(-b/2)], & \text{if } l^{\prime }=l+1, & l \text{ is even}, \\
\gamma _{0}\exp [ik_{x}(b/2)],  & \text{if } l^{\prime }=l+3, & l \text{ is odd}, \\
0,                                            & \text{otherwise}.           &
\end{array}
\right.
\label{eq:ArmchairHmtx}
\end{equation}
For ZGNRs, the Hamiltonian elements are
\begin{equation}
H_{l,l^{\prime }}=\left\{
\begin{array}{ l l l }
2\gamma _{0}\cos (k_{x}a/2), & \text{if } l^{\prime }=l+1, & l \text{ is odd}, \\
\gamma _{0},                         & \text{if } l^{\prime }=l+1, & l \text{ is even}, \\
0,                                          & \text{otherwise},            &
\end{array}
\right.
\label{eq:ZigzagHmtx}
\end{equation}
where $\gamma _{0}=2.598$ eV~\cite{Phys.Rev.B43(1991)4579J.C.Charlier}.
The eigenvalues $E^{h}$'s and eigenstates $|\Psi ^{h}\rangle$'s are obtained by diagonalizing the Hamiltonian matrix.
The superscript $h$ can be either $c$ or $v$ representing the conduction or valence subband.

The energy dispersions of GNRs are symmetric with respect to $k_{x}=0$, and therefore only the positive ones are discussed.
For the AGNR of $N_{y}=69$, many 1D parabolic subbands due to the finite width of GNR are symmetric about the Fermi level [Fig.~\ref{fig:GeometryAndBS}(a)].
The bottoms and tops of these subbands are at $k_{x}=0$, where many band-edge states exist and optical transitions may occur.
The energy gap is equal to $0.05$ $\gamma _{0}$ for this semiconducting nanoribbon.
The subband index $J^{c,v}$ ($=1,2,3,...$) is decided by the minimum energy spacing between the subbands and the Fermi level.
For the ZGNR of $N_{y}=69$, the parabolic subbands are also symmetric about $E_{F}=0$, and their band edges are located at $k_{x}=2\pi /3$ [Fig.~\ref{fig:GeometryAndBS}(b)].
Furthermore, there are degenerate partial flat bands lying on the Fermi level due to the zigzag edge structure, and the dispersionless region extends from $2\pi /3$ to $\pi$.
Hence, it is a gapless semiconductor.

\section{Features of wavefunctions in real space}

\begin{figure}
\begin{center}
  \includegraphics[]{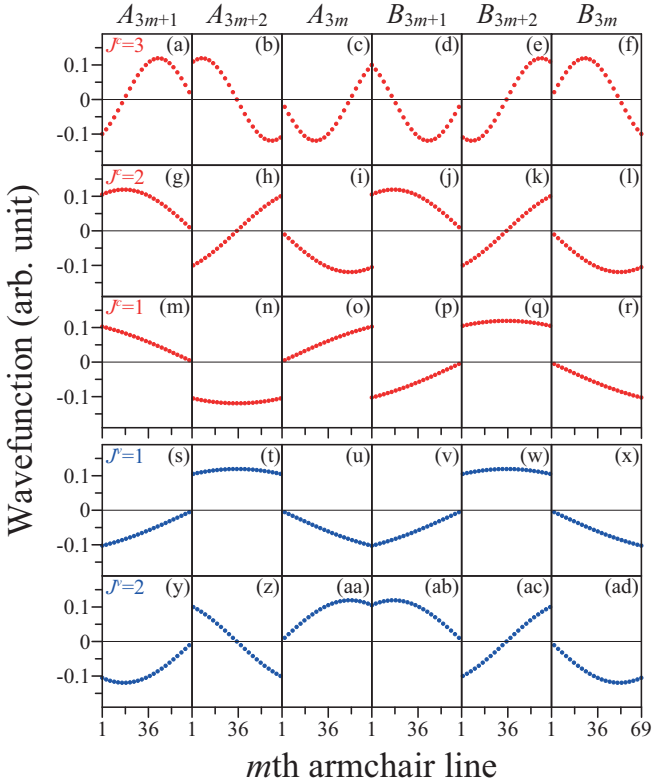}\\
  \caption{
Wavefunctions of subband index $J^{c}=1$--$3$ (red dots) and $J^{v}=1$ and $2$ (blue dots) at $k_{x}=0$ for $N_{y}=69$ AGNR.}
  \label{fig:WFofArmchair}
\end{center}
\end{figure}

Wavefunctions give information on the charge density distribution, and their features are very important in understanding optical selection rules.
They are strongly dependent on the edge structure of nanoribbon.
For AGNRs, the wavefunction is a combination of six subenvelope functions
\begin{widetext}
\begin{equation}
|\Psi ^{h} \rangle = \sum_{m=1,2,3,...}
\bigg[ A_{3m}^{h} |a_{3m}\rangle +A_{3m+1}^{h} |a_{3m+1}\rangle +A_{3m+2}^{h} |a_{3m+2}\rangle
+B_{3m}^{h} |b_{3m}\rangle +B_{3m+1}^{h} |b_{3m+1}\rangle +B_{3m+2}^{h} |b_{3m+2}\rangle \bigg] ,
\label{eq:ArmWFDecomposed}
\end{equation}
\end{widetext}
where $A(B)_{3m}^{h}$'s, $A(B)_{3m+1}^{h}$'s and $A(B)_{3m+2}^{h}$'s, are the site amplitudes of sublattice $A$ ($B$) located at the ($3m$)th, ($3m+1$)th and ($3m+2$)th armchair lines with $m$ being an integer.
The wavefunctions of $N_{y}=69$ AGNR at $k_{x}=0$, where the optical transition occurs, are shown in Fig.~ \ref{fig:WFofArmchair}.
They have an oscillating pattern, and the number of nodes (zero points) is given by $J^{h}-1$.
The square of wavefunction is the charge density distribution.
The local maxima and minima of wavefunctions are of high charge density and the nodes are zero.
As the state energy $|E^{h}|$ increases, the oscillations become more severe.
More importantly, each of them exhibits two aspects of features.
For a certain wavefunction, there is a relation between the subenvelope functions of sublattices $A$ and $B$, that is, the site amplitudes $A_{m}^{h}(J^{h})$ and $B_{m}^{h}(J^{h})$ abide by the relations
\begin{equation}
A_{m}^{h}(J^{h})=\pm B_{m}^{h}(J^{h}).
\label{eq:ArmRelationAhBh}
\end{equation}
For example, the site amplitudes of sublattice $A$ of $J^{c}=1$ subband, $A_{m}^{c}(J^{c}=1)$ [Figs.~ \ref{fig:WFofArmchair}(m)--\ref{fig:WFofArmchair}(o)], is the negative of $B_{m}^{c}(J^{c}=1)$ [Figs.~\ref{fig:WFofArmchair}(p)--\ref{fig:WFofArmchair}(r)].
For the conduction and valence wavefunctions of the same index, $J^{c}=J^{v}=J$, their subenvelope functions have the relations
\begin{equation}
A_{m}^{c}(J)=\pm A_{m}^{v}(J).
\label{eq:ArmRelationAcAv}
\end{equation}
For instance, $A_{m}^{c}(J=2)=-A_{m}^{v}(J=2)$ [compare Figs.~\ref{fig:WFofArmchair}(g)--\ref{fig:WFofArmchair}(i) with Figs.~\ref{fig:WFofArmchair}(y)--\ref{fig:WFofArmchair}(aa)].
The relations between the subenvelope functions of the same subband index are either in phase or out of phase.

\begin{figure}
\begin{center}
  \includegraphics[]{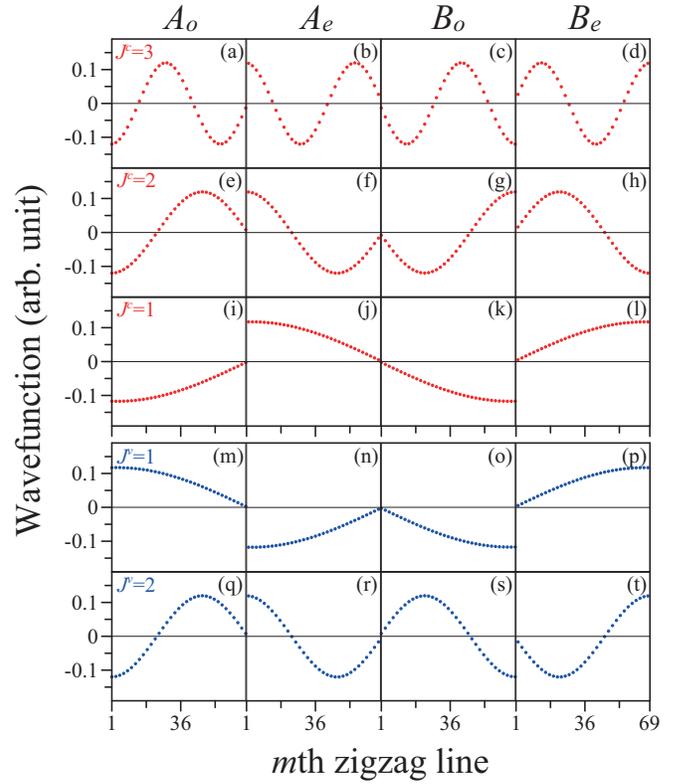}\\
  \caption{
Wavefunctions of subband index $J^{c}=1$--$3$ (red dots) and $J^{v}=1$ and $2$ (blue dots) at $k_{x}=2\pi /3$ for $N_{y}=69$ ZGNR.}
  \label{fig:WFofZigzag}
\end{center}
\end{figure}

For ZGNRs, the wavefunction is decomposed as
\begin{equation}
|\Psi ^{h}\rangle =
  \sum_{odd}A_{o}^{h} |a_{o}\rangle +\sum_{even}A_{e}^{h} |a_{e}\rangle
+\sum_{odd}B_{o}^{h} |b_{o}\rangle +\sum_{even}B_{e}^{h} |b_{e}\rangle ,
\label{eq:ZigWFDecomposed}
\end{equation}
where $A_{o}^{h}$'s, $B_{o}^{h}$'s, $A_{e}^{h}$'s and $B_{e}^{h}$'s are the site amplitudes of sublattices $A$ and $B$ on odd and even zigzag lines.
$|a_{o}\rangle $'s, $|b_{o}\rangle $'s, $|a_{e}\rangle $'s, and $|b_{e}\rangle $'s are the corresponding tight-binding functions.
The wavefunctions of $N_{y}=69$ ZGNR at $k_{x}=2\pi/3$, as shown in Fig.~\ref{fig:WFofZigzag}, reveal profiles different from armchair ones.
The subenvelope functions of the conduction and valence subbands follow the relations:
\begin{equation}
  \begin{array}{l}
    A_{m}^{c}(J^{c})=(-1)^{J^{c}+1}B_{N_{y}+1-m}^{c}(J^{c}), \\
    A_{m}^{v}(J^{v})=(-1)^{J^{v}}B_{N_{y}+1-m}^{v}(J^{v}),
  \end{array}
\label{eq:ZigRelationOdd}
\end{equation}
where the indices ($N_y+1-m$) and $m$ are correlated and the sign factor $(-1)^{J^{c}+1}$ [$(-1)^{J^{v}}$] describes whether the wavefunction is symmetric (positive sign) or anti-symmetric (negative sign).
For example, the subband of $J^{c}=1$ is symmetric [compare Figs.~\ref{fig:WFofZigzag}(i), \ref{fig:WFofZigzag}(j) with Figs.~\ref{fig:WFofZigzag}(k), \ref{fig:WFofZigzag}(l)] and anti-symmetric for $J^{v}=1$ [compare Figs.~\ref{fig:WFofZigzag}(m), \ref{fig:WFofZigzag}(n) with Figs.~\ref{fig:WFofZigzag}(o), \ref{fig:WFofZigzag}(p)].
It is noteworthy that the symmetry property is associated with the subband index.
Considering the conduction subbands, the $J^{c}=1$, $2$, and $3$ wavefunctions are symmetric, anti-symmetric and symmetric, respectively [Figs.~\ref{fig:WFofZigzag}(a)--\ref{fig:WFofZigzag}(l)].
In other words, the symmetry property changes alternatively from symmetry to anti-symmetry with the increment of subband index.
Moreover, the results of this study show that the characteristics of the subenvelope functions $A_{m}^{c}(J^{c})$ and $B_{m}^{c}(J^{c})$ are very sensitive to $N_{y}$ being odd or even.
If $N_{y}$ is even, the relations become
\begin{equation}
  \begin{array}{l}
    A_{m}^{c}(J^{c})=(-1)^{J^{c}}B_{N_{y}+1-m}^{c}(J^{c}), \\
    A_{m}^{v}(J^{v})=(-1)^{J^{v}+1}B_{N_{y}+1-m}^{v}(J^{v}),
  \end{array}
\label{eq:ZigRelationEven}
\end{equation}
which differ from Eq.~(\ref{eq:ZigRelationOdd}) by a minus sign.
However, the alternatively changing symmetry property remains whether $N_{y}$ is odd or even.

\section{Absorption spectra}

\begin{figure}
\begin{center}
  \includegraphics[]{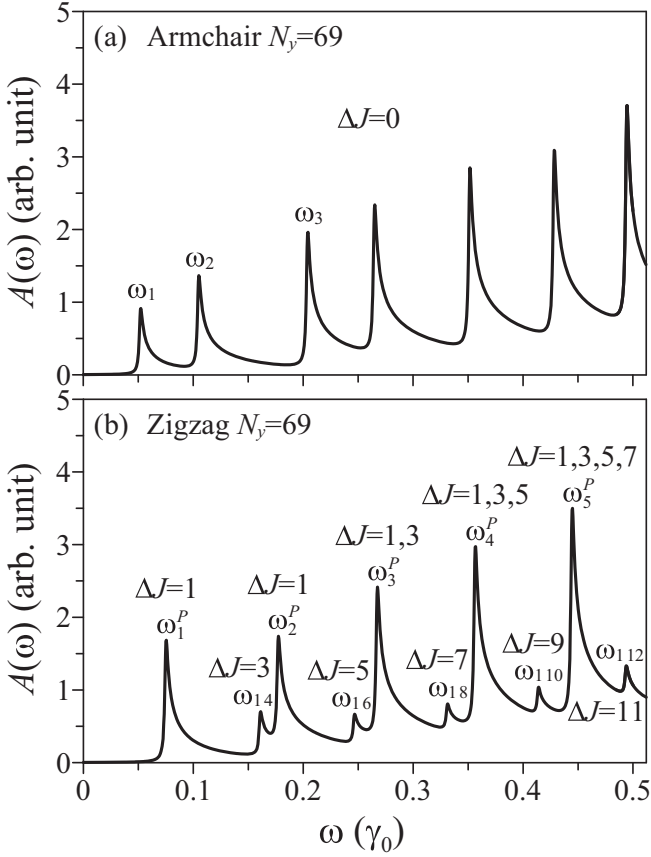}\\
  \caption{
Absorption spectra for $N_{y}=69$ AGNR and ZGNR.}
  \label{fig:AbsSpec}
\end{center}
\end{figure}

When GNR is under the influence of an electromagnetic field with polarization $\mathbf{E} \Vert \hat{x}$, the absorption spectrum $A(\omega)$ for direct photon transitions ($\Delta k_{x}=0$) from state $|\Psi^{h}(k_{x},J^{h})\rangle$ to state $|\Psi^{h'}(k_{x},J^{h'})\rangle$ is given by
\begin{widetext}
\begin{equation}
A(\omega) \propto \sum_{h,h',J^{h},J^{h'}} \int_{1st\text{ B.Z.}} \frac{dk_{x}}{2\pi}
\left\vert \langle\Psi^{h^\prime}(k_{x},J^{h^\prime})|\frac{\hat{\mathbf{E}}\cdot\mathbf{P}}{m_e}|\Psi^{h}(k_{x},J^{h})\rangle\right\vert ^2
\times \mathrm{Im} \left\{ \frac{f[E^{h}(k_{x},J^{h})]-f[E^{h^\prime}(k_{x},J^{h^\prime})]}{E^{h^\prime}(k_{x},J^{h^\prime})-E^{h}(k_{x},J^{h})-\omega -i\Gamma }\right\},
\label{eq:AbsFormula}
\end{equation}
\end{widetext}
where $m_{e}$ is the effective mass of electron, $\mathbf{P}$ is the momentum operator, $f[E^{h}(k_{x},J^{h})]$ is the Fermi-Dirac distribution function, and $\Gamma$ is the broadening factor.
The low-energy absorption spectra for $N_{y}=69$ AGNR and ZGNR at zero temperature are illustrated in Fig.~ \ref{fig:AbsSpec}.
The spectra show a lot of 1D peaks caused by the inter-band transitions of 1D subbands, and they are greatly affected by the geometric structure.

For absorption spectrum of the $N_{y}=69$ AGNR [Fig.~\ref{fig:AbsSpec}(a)], the first peak, located at $\omega_{1}=0.055$ $\gamma _{0}$, is numerically identified as the excitation from $J^{v}=1$ to $J^{c}=1$ subbands.
The second ($\omega _{2}=0.105$ $\gamma _{0}$) and the third ($\omega _{3}=0.205$ $\gamma _{0}$) ones are the transitions from $J^{v}=2$ to $J^{c}=2$ subbands and from $J^{v}=3$ to $J^{c}=3$ subbands, respectively.
The allowed transitions originate in the subbands of the same indices, i.e., the selection rule is $\Delta J=J^{c}-J^{v}=0$.
The transition energy is twice the band-edge energy ($\omega_J=2E^{edge}_{J^c}=2|E^{edge}_{J^v}|$, where $J^c=J^v=J$) because the band structure is symmetric about $E_{F}=0$.
Moreover, the peak arises higher when the frequency increases as a consequence of the decrease of subbands' curvatures.

For ZGNRs, the absorption peaks may be classified into principal peaks and subpeaks according to their heights, as shown in Fig.~\ref{fig:AbsSpec}(b).
The allowed transitions obey the selection rule $|\Delta J|=odd$, which is different from the armchair case, $\Delta J =0$.
The principal peaks become higher with the increasing frequency because of the decrease of bands' curvatures and the increase of excitation channels.
Due to the symmetry of energy spectrum, the absorption spectrum of negative $\Delta J$'s is the same as that of positive ones.
Therefore, only the positive $\Delta J$ cases (transitions from $J^{v}$ to $J^{c}=J^{v}+\Delta J $ subband of energy $\omega=E^{edge}_{J^v+\Delta J}-E^{edge}_{J^v}$) are discussed.
The principal peaks, $\omega^P_J$, are contributed by transitions between subbands satisfying the selection rule $\Delta J=1$ $(J^v\rightarrow J^c: 1\rightarrow 2)$, $\Delta J=1$  $(J^v\rightarrow J^c: 2\rightarrow 3)$, $\Delta J=1,3$  $(J^v\rightarrow J^c:3 \rightarrow 4,~2\rightarrow 5,)$, $\Delta J=1,3,5$ $(J^v\rightarrow J^c:4 \rightarrow 5,~ 3\rightarrow6,~2\rightarrow 7)$, and so on.
The corresponding numbers of excitation channels are 2, 2, 4, and 6, i.e., the double of the allowed positive $\Delta J$ number.
On the other hand, there exists one subpeak between two adjacent principal peaks.
These subpeaks are indicated by $\omega_{1J}$, where the subscript $1J$ denotes the transition from the $J^v=1$ valance subband to the $J^c$ conduction subband.
The subpeak $\omega_{14}$, for instance, results from the transition between the $J^v=1$ and  $J^c=4$ subbands and follows the optical selection rule $\Delta J=3$.
Since each subpeak possesses two excitation channels, the increase of subpeak height is not as fast as the principal one.
It is remarkable that only the first principal peak ($\omega^P_1$) and all the subpeaks ($\omega_{1J}$) are from the transitions related to the $J^{c,v}=1$ subbands ($E_{J^{c,v}=1}=E_F=0$).
The $\omega^P_1$ peak is located at $E^{edge}_{J^c=2}$, and the subpeaks $\omega_{1J}$ are positioned at $E^{edge}_{J^c=J}$, which is the energy difference between the subband edge and the Fermi level.

\section{Optical selection rules of GNRs}

In addition to the above-discussed features of wavefunctions, the optical selection rules are also determined by the velocity matrix elements [$\mathbb{M}^{hh'}(k_x) \equiv \langle \Psi ^{h^{\prime }}(k_{x},J^{h^{\prime }})| (\hat{\mathbf{E}} \cdot \mathbf{P} / m_e) |\Psi ^{h}(k_{x},J^{h})\rangle$].
To evaluate $\mathbb{M}$, the gradient approximation ($\hat{\mathbf{E}}\cdot\mathbf{P}/m_{e}=\mathbf{\nabla }_{k}H$) is employed~\cite{Phys.Rev.B7(1973)2275L.G.Johnson}, so that
\begin{equation}
\mathbb{M}^{hh'}(k_x) = \sum_{l,l^{\prime }=1}^{2N_{y}}C_{l}^{h\ast }(k_x)C_{l^{\prime}}^{h^{\prime }}(k_x) \frac{\partial H_{l,l^{\prime }}(k_x)}{\partial k_{x}} .
\label{eq:velocity_matrix_element_1}
\end{equation}

At zero temperature, only the inter-band transitions from the valence to conduction subbands are valid.
To investigate the optical properties, we focus on $k_{x}=0$, where the excitation occurs for AGNRs.
Substituting the sublattices B with A in Eq.~(\ref{eq:velocity_matrix_element_1}) by the relations $B_{m}^{v}=sA_{m}^{v}$ ($s=\pm 1$) and $B_{m}^{c}=tA_{m}^{c}$ ($t=\pm 1$) from Eq.~(\ref{eq:ArmRelationAhBh}), $\mathbb{M}$ is
\begin{eqnarray}
\nonumber 
b\gamma_{0}  \sum\limits_{m=1}^{N_{y}} && \bigg[ -sA_{m}^{c\ast}(J^{c})A_{m}^{v}(J^{v})+tA_{m}^{c\ast }(J^{c})A_{m}^{v}(J^{v}) \bigg]
\\
\nonumber 
+ \frac{b\gamma_0}{2} \sum\limits_{m=1}^{N_{y}-1} && \bigg[ -tA_{m}^{c\ast}(J^{c})A_{m+1}^{v}(J^{v})+sA_{m}^{c\ast }(J^{c})A_{m+1}^{v}(J^{v}) \bigg]
\\
\nonumber
+ \frac{b\gamma_0}{2} \sum\limits_{m=2}^{N_{y}} && \bigg[ -tA_{m}^{c\ast}(J^{c})A_{m-1}^{v}(J^{v})+sA_{m}^{c\ast }(J^{c})A_{m-1}^{v}(J^{v}) \bigg] .
\\
\label{eq:Before Using H}
\end{eqnarray}
The transfer relations between $A_{m\pm 1}$ and $A_{m}$, obtained from $H|\Psi\rangle=E|\Psi\rangle$ at $k_{x}=0$, are
\begin{equation}
\left\{
\begin{array}{lc}
A_{2}=\Delta _{\pm }A_{1},                     &  \\
A_{m-1}+A_{m+1}=\Delta _{\pm }A_{m}, & \text{if } m=2,3,4,...,N_{y}-1, \\
A_{N_{y}-1}=\Delta _{\pm }A_{N_{y}},     &
\end{array}
\right.
\label{eq:Am pm 1 to Am}
\end{equation}
where $\Delta_{\pm }=\pm (E/\gamma _{0})-1$ for $A_{m}=\pm B_{m}$.
After using Eq.~(\ref{eq:Am pm 1 to Am}) to replace $A_{m\pm 1}$ with $A_{m}$, $\mathbb{M}$ becomes
\begin{eqnarray}
\nonumber 
b\gamma _{0} \sum\limits_{m=1}^{N_{y}} \bigg[ -sA_{m}^{c\ast}(J^{c})A_{m}^{v}(J^{v})+tA_{m}^{c\ast }(J^{c})A_{m}^{v}(J^{v})
\\
\nonumber 
-\frac{\Delta _{s}}{2}tA_{m}^{c\ast }(J^{c})A_{m}^{v}(J^{v})+\frac{\Delta _{s}}{2}sA_{m}^{c\ast }(J^{c})A_{m}^{v}(J^{v}) \bigg]
\\
= (t-s)(1-\frac{\Delta _{s}}{2})b\gamma_{0}\sum\limits_{m=1}^{N_{y}}A_{m}^{c\ast }(J^{c})A_{m}^{v}(J^{v}),
\end{eqnarray}
where $\Delta_{s}=s(E/\gamma _{0})-1$ for $A_{m}^{c}=sB_{m}^{c}$.
Through the relation between the conduction and valence subbands $A_{m}^{v}=uA_{m}^{c}$ ($u=\pm 1$) from Eq.~ (\ref{eq:ArmRelationAcAv}) and the normalization of the wavefunction $2\sum_{m=1}^{N_{y}}A_{m}^{c\ast}(J^{c})A_{m}^{c}(J^{v})=\delta _{J^{c},J^{v}}$, the velocity matrix element at $k_{x}=0$ is
\begin{eqnarray}
\nonumber 
\mathbb{M}^{vc}(0) & = & u(t-s)(1-\frac{\Delta _{s}}{2})b\gamma _{0}\frac{1}{2}\delta_{J^{c},J^{v}}
\\
& = & \pm (1-\frac{\Delta _{\pm }}{2})b\gamma _{0}\delta_{J^{c},J^{v}} ,
\label{eq:Arm selection rule}
\end{eqnarray}
where $\Delta _{\pm }=\pm (E/\gamma _{0})-1$ and $u(t-s)$ is either $0 $ or $\pm 2$.
Eq.~(\ref{eq:Arm selection rule}) achieves the selection rule $(\Delta J=0$) for AGNRs and possible excitation channels in absorption spectra.

For ZGNRs, the velocity matrix element in Eq.~(\ref{eq:velocity_matrix_element_1}) at $k_{x}=2\pi /3$ becomes
\begin{equation}
h^{\prime }\sum_{m=1}^{N_{y}} \bigg[ A_{m}^{c\ast}(J^{c})B_{m}^{v}(J^{v})+B_{m}^{c\ast }(J^{c})A_{m}^{v}(J^{v}) \bigg],
\label{eq:WF_Zig_1}
\end{equation}
where $h^{\prime }=-a\gamma _{0}\sin (\pi a/3)$.
For odd $N_{y}$, we use Eq.~(\ref{eq:ZigRelationOdd}) to convert the second term in Eq.~(\ref{eq:WF_Zig_1}), which leads to
\begin{eqnarray}
\nonumber 
h^{\prime } \sum_{m=1}^{N_{y}} && \bigg[ A_{m}^{c\ast}(J^{c})B_{m}^{v}(J^{v})
\\
+(-1&&)^{J^{c}+1}A_{N_{y}+1-m}^{c\ast}(J^{c})(-1)^{J^{v}}B_{N_{y}+1-m}^{v}(J^{v}) \bigg].
\label{eq:WF_Zig_2}
\end{eqnarray}
Then renaming the index of $N_{y}+1-m$ as $m$ gives
\begin{equation}
[1+(-1)^{J^{c}+J^{v}+1}]h^{\prime}\sum_{m=1}^{N_{y}}A_{m}^{c\ast}(J^{c})B_{m}^{v}(J^{v}).
\label{eq:Zig selection rule Ny odd}
\end{equation}
As for even $N_{y}$, the second term in Eq.~(\ref{eq:WF_Zig_2}) can also be converted by Eq.~(\ref{eq:ZigRelationEven}), and it yields the same result as Eq.~(\ref{eq:Zig selection rule Ny odd}).
Hence, $\mathbb{M}^{vc}$ for ZGNR at $k_{x}=2\pi /3$ is
\begin{equation}
\mathbb{M}^{vc}(\frac{2\pi}{3}) = \left\{
\begin{array}{ll}
2h' \sum_{m=1}^{N_{y}}A_{m}^{c\ast}(J^{c})B_{m}^{v}(J^{v}) , & \text{if } \Delta J=odd, \\
0,                                                                                         & \text{if } \Delta J=even,
\end{array}
\right.
\label{eq:Zig selection rule}
\end{equation}
where $\sum_{m=1}^{N_{y}}A_{m}^{c\ast }(J^{c})B_{m}^{v}(J^{v})$ has a non-zero value, and the selection rule ($\Delta J=odd$) for ZGNRs is obtained.

\begin{figure}
\begin{center}
  \includegraphics[]{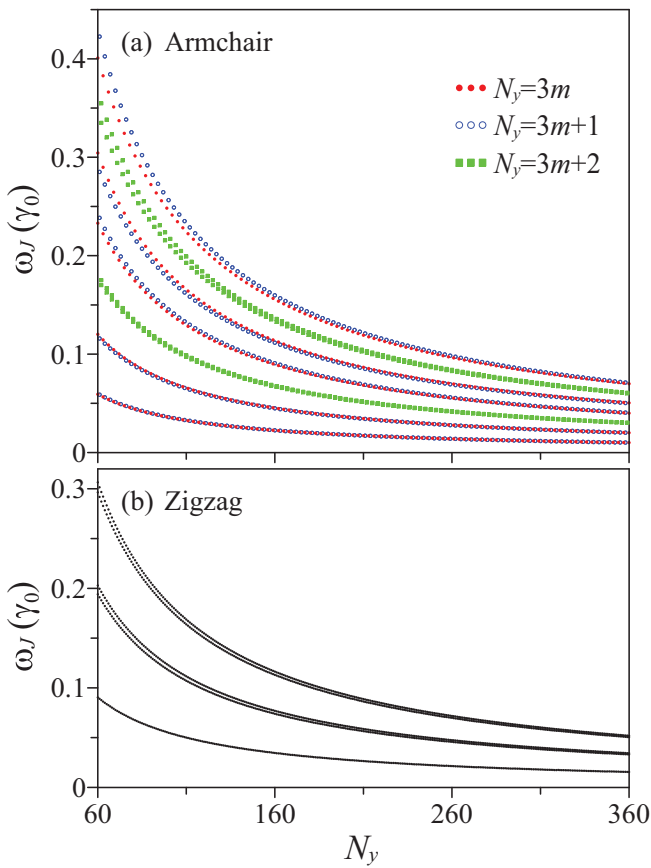}\\
  \caption{
The first five consecutive transition energies with respect to the ribbon width $N_{y} $ for AGNR and ZGNR.
The notations $\bullet$, $\circ$ and $\blacksquare$ correspond to the AGNRs of $N_{y}=3m$, $3m+1$, and $3m+2$, respectively.}
  \label{fig:ETvsNy}
\end{center}
\end{figure}

By using the selection rules, the transition energies (peak positions) can be efficiently obtained from the energy dispersion.
For AGNRs, the dependence of the first five predicted transition energies on the ribbon width is shown in Fig.~ \ref{fig:ETvsNy}(a).
For a certain width $N_y$, the $J$th transition energy is equal to $\omega_J=2E^{edge}_{J^c}$, i.e., twice of the band-edge state energy of the $J^c$th subband.
For narrower GNRs, there are three groups of transition energies.
The $\omega_J$'s of $N_{y}=3m$ (full circles) and $N_{y}=3m+1$ (open circles) groups are close.
While for the $N_{y}=3m+2$ (squares) group, the first and second $\omega_J$'s are close and sandwiched between the second and third $\omega_J$'s of the $N_{y}\neq 3m+2$ nanoribbons.
For wider GNRs, these transition energies make red-shift and merge together.
The GNRs may therefore be sorted into two groups of $N_{y}= 3m+2$ and $N_{y}\neq 3m+2$.
The former is gapless due to the linear bands crossing at the Fermi level, while the latter is semiconducting with a band gap energy corresponding to the first transition energy.

The first five ribbon-width-dependent transition energies of ZGNRs are presented in Fig.~\ref{fig:ETvsNy}(b).
They are associated with the first principal peak $\omega^P_1$, the first subpeak $\omega_{14}$, the second principal peak $\omega^P_2$, the second subpeak $\omega_{16}$, and the third principal peak $\omega^P_3$, respectively [see Fig.~ \ref{fig:AbsSpec}(b)].
The predicted corresponding transition energies are
$\omega^P_1 = E^{edge}_{J^c=2}$,
$\omega_{14} = E^{edge}_{J^c=4}$,
$\omega^P_2 = E^{edge}_{J^c=3} - E^{edge}_{J^v=2}$,
$\omega_{16} = E^{edge}_{J^c=6}$, and
$\omega^P_3 = E^{edge}_{J^c=4} - E^{edge}_{J^v=3}$.
The first subpeak and the second principal peak are close and will merge for sufficiently large width, so do the second subpeak and the third principal peak.
In short, the peak frequencies can be predicted by the combination of band structures and selection rules.
This is an efficient way to obtain wide-range information on the transition energies without extrapolation.

It is worth mentioning that the exact energies of band-edge states for ZGNRs can be specified by the experimentally measured transition energy $\omega^{exp}_{J}$.
According to the optical selection rule, the measured peak positions $\omega^{P,exp}_1$, $\omega^{exp}_{14}$, and $\omega^{exp}_{16}$ can be applied to specify the band-edge states $E^{edge}_{J^c=2}$, $E^{edge}_{J^c=4}$, and  $E^{edge}_{J^c=6}$, respectively.
They are the energy differences between the $J^c=even$ subband edges and the Fermi level.
Moreover, the band-edge state energies of the $J^c=odd$ subbands are obtainable, for example, $E^{edge}_{J^c=3}=\omega^{P,exp}_2 - |E^{edge}_{J^v=2}| = \omega^{P,exp}_2 - \omega^{P,exp}_1$.
Similar backtracking strategy is valid for AGNRs based on the band symmetry about the Fermi energy.
The band-edge state energy, which is the half of the experimentally measured peak position, is given by $E_{J^c}^{edge}=|E_{J^v}^{edge}|=\omega^{exp}_{J}/2$, where $J^c=J^v=J$.
Hence, the selection rule provides a way to overcome the disadvantage of optical measurement, that is, it only yields the information about the energy difference between two subbands.

\section{Conclusion}

We employ the tight-binding model to study the absorption spectra of GNRs, and this is for the first time that the corresponding optical selection rules are analytically specified.
The main results of this work are stated as follows.
First, the optical transition channels of absorption spectra for AGNRs and ZGNRs are exactly identified.
Then, the characteristics of absorption peaks, such as positions and heights, related to the energy dispersions are discussed in detail.
Most importantly, this work provides a theoretical explanation on the optical selection rules through analyzing the velocity matrix elements and the wavefunction features.
The optical selection rules depend on the edge structure, armchair or zigzag edges.
The selection rule of AGNRs ($\Delta J=0$) originates from two aspects: the relation between sublattices $A$ and $B$ within a certain state and the relation between the conduction and valence subenvelope functions of the same index.
While for ZGNRs, the selection rule ($|\Delta J|=odd$) owes to the alternatively changing symmetry property with the increasing index.
According to the selection rules, we can efficiently predict the peak positions and identify the corresponding transition channels.
Furthermore, this gives a way to overcome the limitation of optical measurements.
Hence, the energies of band-edge states can be exactly obtained from optical experiments.


\begin{acknowledgments}
This work was supported in part by the National Science Council of Taiwan under grant numbers NSC 99-2112-M-165-001-MY3 and 98-2112-M-006-013-MY4.
\end{acknowledgments}


%

\end{document}